\documentclass[a4paper,12pt]{revtex4-1}
\usepackage{graphicx}
\usepackage[utf8]{inputenc}
\usepackage{amsmath}
\usepackage{float}
\usepackage{fancyhdr}
\usepackage{hyperref}
\begin{document}

\title{Emulating tightly bound electrons in crystalline solids using mechanical waves}

\author{F. Ram\'{\i}rez-Ram\'{\i}rez}
\affiliation{Posgrado en Ciencias e Ingenier\'ia, Divisi\'on de Ciencias B\'asicas e Ingenier\'ia, Universidad Aut\'onoma Metropolitana-Azcapotzalco, Av. San Pablo 180, Col. Reynosa Tamaulipas, 02200 Ciudad de M\'exico, Mexico}

\author{E. Flores-Olmedo} 
\affiliation{Departamento de Ciencias B\'asicas, Universidad Aut\'onoma Metropolitana-Azcapotzalco, Av. San Pablo 180, Col. Reynosa Tamaulipas, 02200 Ciudad de M\'exico, Mexico}

\author{G. B\'aez} 
\affiliation{Departamento de Ciencias B\'asicas, Universidad Aut\'onoma Metropolitana-Azcapotzalco, Av. San Pablo 180, Col. Reynosa Tamaulipas, 02200 Ciudad de M\'exico, Mexico}

\author{E. Sadurn\'{\i}}
\affiliation{ Instituto de F\'isica, Benem\'erita Universidad Aut\'onoma de Puebla, Apartado Postal J-48, 72570 Puebla, M\'exico}

\author{R.~A. M\'endez-S\'anchez}                           
\affiliation{Instituto de Ciencias F\'isicas, Universidad Nacional Aut\'onoma de
M\'exico, Apartado Postal 48-3, 62210 Cuernavaca Mor., Mexico}

\begin{abstract}
Solid state physics deals with systems composed of atoms with strongly bound electrons. 
The tunneling probability of each electron is determined by interactions that typically extend to neighboring sites, as their corresponding wave amplitudes decay rapidly away from an isolated atomic core. 
This kind of description is essential to material science, and it rules the electronic transport properties of metals, insulators and other condensed matter systems. 
The corresponding phenomenology is well captured by tight-binding models, where the electronic band structure emerges from atomic orbitals of isolated atoms plus their coupling to neighboring sites in a cristal.
In this work, a mechanical system that emulates dynamically a tightly bound electron is built.
This is done by connecting mechanical resonators via locally periodic aluminum bars acting as couplers. 
When the frequency of a particular resonator lies within the frequency gap of a coupler, the vibrational wave amplitude imitates a bound electron orbital. The localization of the wave at the resonator site and its exponential decay along the coupler are experimentally verified.
The quantum dynamical tight-binding model and frequency measurements in mechanical structures show an excellent agreement.

\end{abstract}

\pacs{03.65.-w; 03.65.Xp; 05.60.Gg; 71.20.-b; 71.23.An; 71.55.-i; 71.70.-d; 72.10.Fk; 72.15.Rn; 43.40.+s; 46.70.De; 81.05.Xj; 81.05.Zx}

\maketitle

\section{Main}
\label{Main}
The determination of the electronic band structure in crystalline materials is one of the most important problems in solid-state physics. Fortunately, in many cases, the electrons of a crystal are strongly attached to the atoms in the grid and consequently the band structure can be calculated easily. 
This fact is captured by the tight-binding (TB) model in which a very weak interaction with the neighboring atoms is supposed \cite{Kittel, AshcroftMermin}. 
In the simplest picture of solids, the electronic wave functions of interacting atoms are expanded in terms of wave functions of isolated atoms, i.e. individual atomic orbitals. When the atomic nuclei are located at the sites of a periodic mesh, the corresponding expansion coefficients are given by discrete plane waves of quasi-momentum $k$, in compliance with Bloch’s theorem. 
It is important to note that this expansion obeys exclusively the symmetry of the atomic array, and does not incorporate any detail of the coupling between sites. However, when the interaction sets in, only the plane wave expansion above can be associated to a well-defined energy state: it is delocalized throughout the array (the electron does not belong to any particular site) and it determines the energy band structure by means of a dispersion relation $E(k)$. 
The nearest-neighbor TB is then a particular case when the electron can ``hop'' only between nearest sites and the emergent bandwidth is proportional to the hopping energy or coupling.  
This model gives good quantitative results in many cases and can be combined with other models to improve the results when the TB model is not satisfactory. Interactions with second and higher-order neighbours can also be included. 
The TB model offers the possibility of understanding metals, insulators, magnets and superconductors \cite{GoringeBowlerHernandez}.
This model has also been considered as an ideal platform to explore emergent properties of novel materials, such as graphene \cite{NetoGuineaPeresNovoselovGeim, MendezJureidiniNaumis}, hexagonal boron nitride \cite{ZhaoZhaoWangFan}, stanene, germanene, silicene, among many others \cite{Hattori}. 
Furthermore it can be used to study molecules \cite{Pastawski}.
In addition, the TB model has been applied to study the properties of photonic crystals \cite{BayindirTemelkuranOzbay, Yablonovitch}, 
phononic crystals \cite{MattarelliSecchiMontagna, SainidouStefanouModinos, KushwahaHaleviDobrzynskiDjafari, SigalasEconomou, MundayBradRobertson, Khelif}, 2D electron gasses \cite{Singha} and Bose-Einstein condensates in optical lattices \cite{KraemeMenottiPitaevskiiSandro}. Moreover, the TB model has been emulated in top-table experiments with microwaves, either with resonators mimicking atomic orbitals \cite{PoliBellecKuhlMortessagneSchomerus} or with evanescent modes in waveguides \cite{Bittner, RiverMocinosSadurni}.
Contrary to these cases, up to now, the area of structured elastic systems has remained open for exploration and application of the TB model. 
The main difficulty in the study of this area is that the typical coupling between connected vibrating solids is very strong, and has a long range. 
In this paper, the emulation of evanescent (or weak) couplings via locally periodic structures is presented.
The idea is to use the gaps of crystalline mechanical structures that will be taken as couplers. 
As it is well known, a periodic system, even an elastic one, shows bands and gaps. 
Therefore connecting a resonator with another one through a locally periodic structure, the resonators will communicate weakly one to each other through the coupler when its resonant frequency lies within the coupler's badgap.
The maximum wave amplitude will be located at the position of the resonator and from there, the amplitude will decay exponentially through the coupler \cite{MoralesMendezFlores, RamirezRamirezF}. 
This result is used in order to build a mechanical material with transport through such {\em trapped states}.

\begin{figure}[htb]
\centering
\includegraphics[width=\columnwidth]{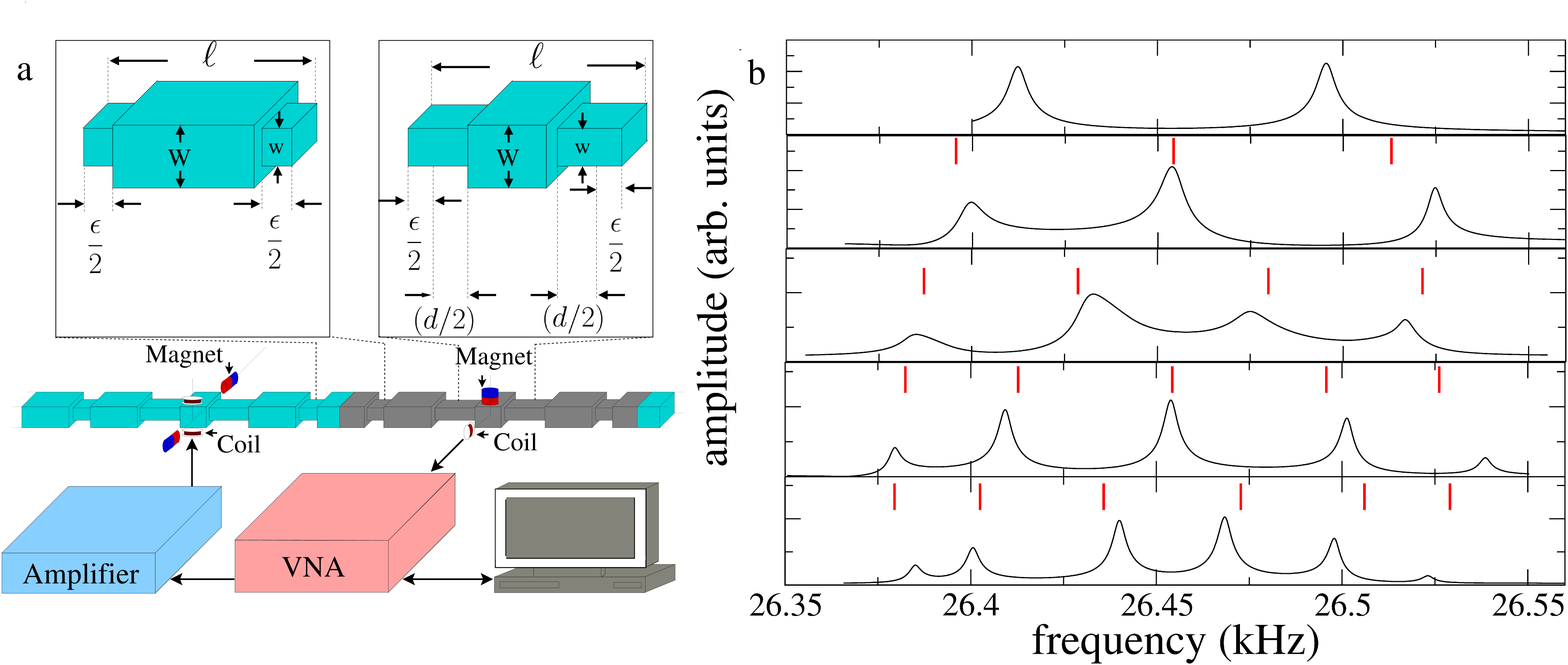}
\caption{\textbf{Schematic representation of the system constructed and the distribution of the frequency levels.}
\textbf{a}, Bottom: Setup used to characterize the vibrations of the elastic structures with acoustic resonant spectroscopy. The setup is composed by a workstation, a vector network analyzer (Anritsu MB-4630B), a high-fidelity audio amplifier (Cerwin-Vega CV-2800), and two electromagnetic-acoustic transducers. 
Middle: the elastic structure to be characterized; it corresponds to two coupled resonators. 
The shadow zone indicate a supercell. 
Top left: detailed view of the unit cell of the coupler; it is composed by three cuboids: the central one with length $\ell-\epsilon$=92~mm, width and height $W$=12.7~mm and two identical small cuboids, of length $\epsilon /2$=4~mm width and height $w=8.7$~mm, at both ends. Top right: a resonating cell, composed by three cuboids, is shown. The central cuboid has length $\ell-\epsilon-d$=55.2~mm, width and height $W$=12.7~mm and two identical small cuboids, of length $(d+\epsilon)/2$=22.4~mm width and height $w=8.7$~mm, at both ends. 
\textbf{b}, From top to bottom measured spectra of the emergent band for the beams with 2, 3, 4, 5 and 6 supercells. The results of the tight-binding model are  indicated by the vertical (red) lines. 
}
\label{zzzUltimo31abcd}
\end{figure}

In this work five mechanical vibrating systems, that obey the \textit{quantum} tight-binding model and emulate a finite {\em quantum} 1D crystal, are reported.
The elastic systems were constructed on aluminum beams and are composed of $n$ resonators joined by couplers (see Fig.~\ref{zzzUltimo31abcd}). 
Each coupler is formed by $m$ unit cells of length $\ell$ and each coupler unit cell is composed of a large cuboid of cross-sectional area $W\times W$ and length $\ell-\epsilon$ and two small cuboids of area $w\times w$ and length $\epsilon/2$ with $W, w, \epsilon \ll \ell$ (see Fig.~\ref{zzzUltimo31abcd} {\bf a}, left inset). 
The resonator also has length $\ell$, and is composed of one cuboid of $W\times W$ and length $l-\epsilon-d$ and two small cuboids of  $w\times w$ and length $(d + \epsilon)/2$ (see right inset of Fig.~\ref{zzzUltimo31abcd} {\bf a}). Two couplers of $M$($>m$) cells are used as terminators to avoid finite size coupler's border effects. 
The vibrating system can also be understood as a crystal composed by $n$ supercells (see shadow zone in the beam of Fig.~\ref{zzzUltimo31abcd}). 
Each system was constructed by machining a solid aluminum piece.

The  experimental spectra of the structured rods, for torsional waves, corresponding to two up to six {\em artificial elastic atoms} or supercells are shown in Fig.~\ref{zzzUltimo31abcd} b.
These spectra were measured using acoustic resonant spectroscopy exciting at the position of one resonator and detecting at the position of other resonator.
In this figure the spectra show the emergence of a band, located within the second bandgap of the coupler, as the number of supercells increases.
As it can be seen the band emerges symmetrically approximately at 26450~Hz, the frequency of a single artificial elastic atom. 
Each band is formed by well resolved Breit-Wigner line shapes with a number of resonances given by the number of artificial elastic atoms. 
The width of the band is approximately 150~Hz and it is possible to notice that the level spacings at the borders of the band are smaller than the level spacings at the center of the band, in agreement with the energy spectrum of a 1D atomic crystal.

\begin{figure}[htb]
\centering
\includegraphics[width=\columnwidth]{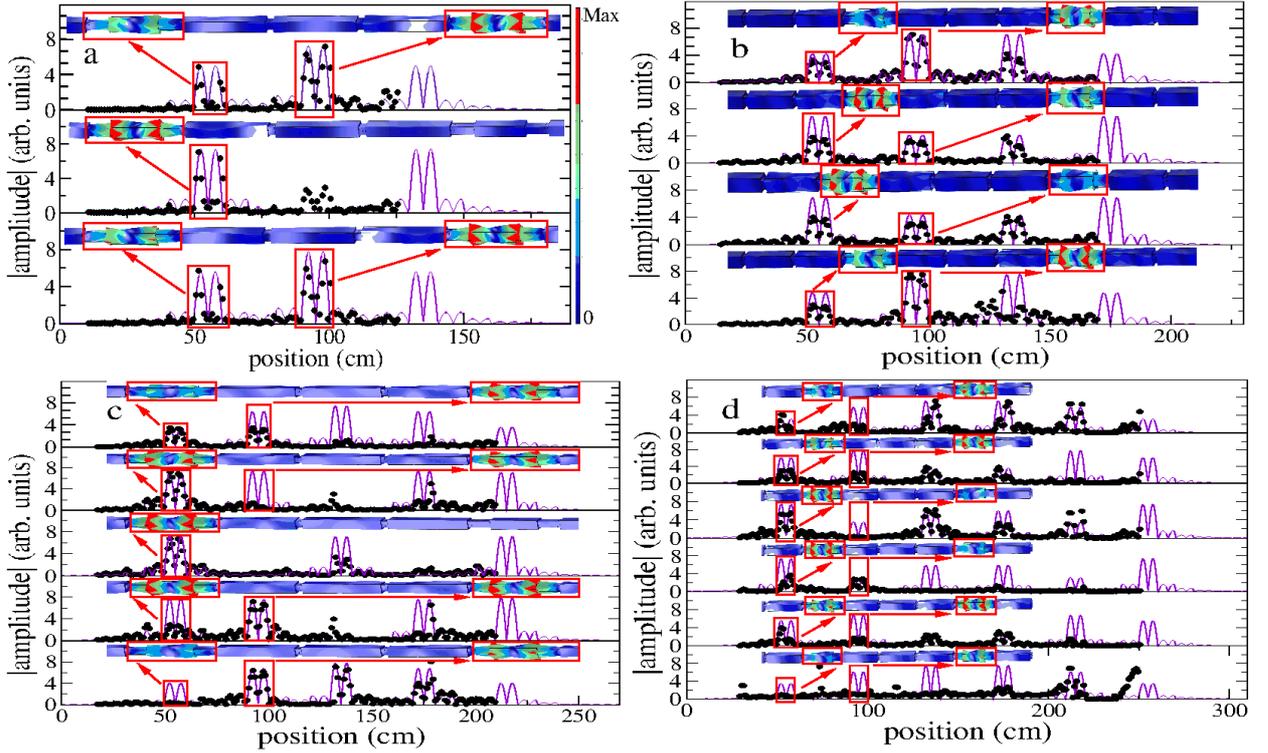}
\caption{{\textbf{Experimental vs. FEM  vs. TB model wave amplitudes.}
The absolute values of the torsional wave amplitudes, as function of the position, for the structured system with three, four, five and six coupled resonators are given in {\bf a},  {\bf b},  {\bf c} and  {\bf d}, respectively. The experimental and TB model results are given by dots and the continuous lines, respectively. 
In the upper part of each plot an amplification of the deformations of the elastic structure, obtained with finite elements, around two consecutive resonators, is shown.
The color scale show the maxima (minima) of the deformations in red (blue). 
 In descending order the amplitudes correspond to levels in the emergent band with frequencies 
\textbf{a}, $f_{1}$ = 26472~Hz, $f_{2}$= 26544~Hz and $f_{3}$ = 26613~Hz. 
\textbf{b}, $f_{1}$ = 26461~Hz, $f_{2}$= 26512~Hz, $f_{3}$=26574~Hz and $f_{4}$ =26622~Hz.
\textbf{c}, $f_{1}$=  26455~Hz, $f_{2}$=26492~Hz, $f_{3}$=26543~Hz, $f_{4}$=26592~Hz  and $f_{5}$=26628~Hz, and
\textbf{d}, $f_{1}$= 26451~Hz, $f_{2}$= 26480~Hz, $f_{3}$= 26520~Hz, $f_{4}$= 26565~Hz, $f_{5}$= 26604~Hz and $f_{6}$= 26631~Hz, respectively.
}}
\label{Figura5}
\end{figure}

In  Fig.~\ref{Figura5}, top left (top right, bottom left, bottom right) the measurements of the absolute value of the torsional wave amplitudes, as a function of the position, for the elastic structure with three (four, five, six) coupled resonators are shown. 
Each wave amplitude has a one-to-one correspondence with each frequency level in the emergent band of Fig.~\ref{zzzUltimo31abcd}. 
The obtained data were recorded by moving the detector along the beam; only two thirds of the total length of the structure were measured to avoid the saturation of the detector by the exciter's magnetic field. 
Note that the experimental wave amplitudes are localized at the position of the resonators and that they show an exponential decay in the couplers.
This is characteristic of states associated to frequencies in the bandgap and are compatible with the case of electrons  strongly bounded to their atoms (tight-binding electrons). 
This fact is in agreement with the wave amplitudes obtained from the finite element numerical calculation in which the deformations of the structured beam can be seen mainly localized at the resonator positions.

\begin{figure}[htb]
\centering
\includegraphics[width=\columnwidth]{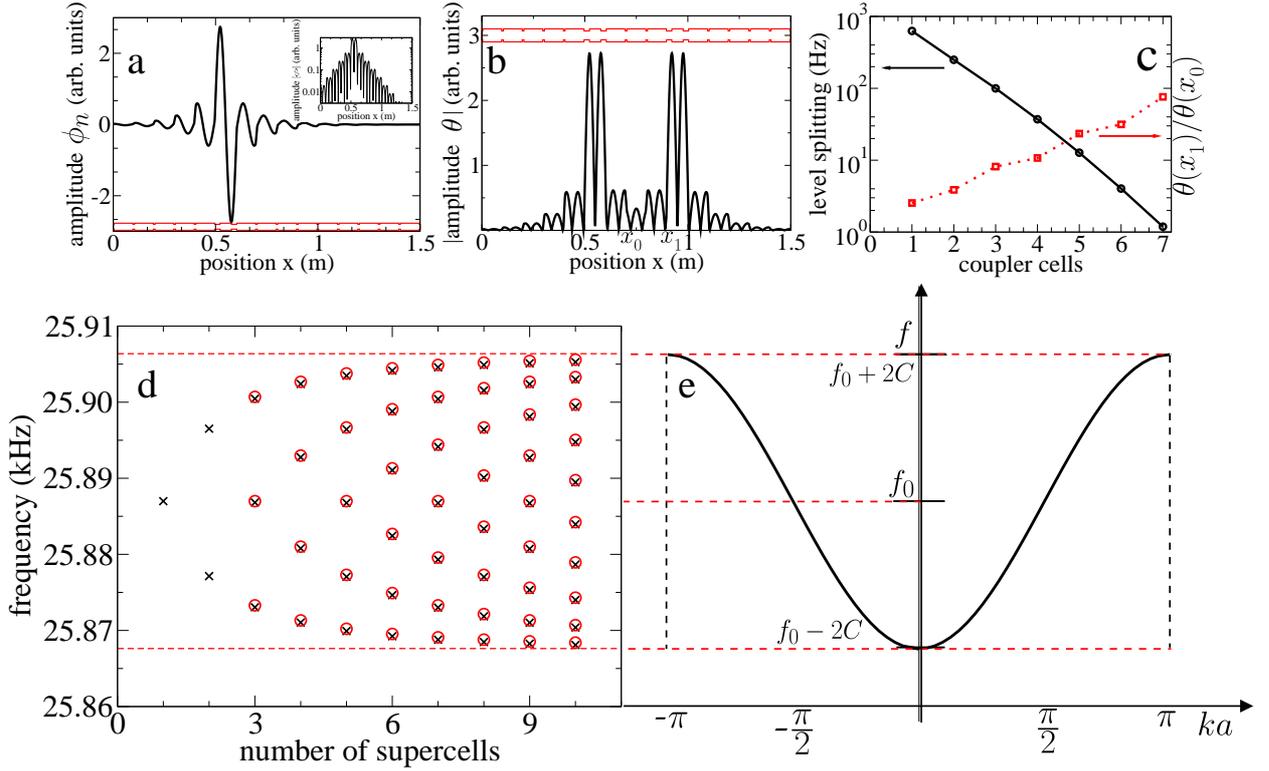}
\caption{
\textbf{Wave amplitudes, level repulsion and dispersion relation of the emergent crystal.} 
\textbf{a}, Torsional wave amplitude $\phi_n(x)$, obtained with the transfer matrix method, as function of the position for a locally periodic beam with a resonator on site $n$. 
The longitudinal section of the elastic structure is shown in the lower part (red color). 
The exponential decay is shown in the inset. 
\textbf{b}, Numerical wave amplitude $\theta(x)$ (absolute value) as a function of the position obtained with the transfer matrix method, for a a locally periodic beam with two coupled resonator. 
\textbf{c}, Vertical left axis: level spacing as a function of the number of cells of the coupler.  
Vertical right axis: ratio $\theta(x_1)/\theta(x_0)$ where $x_0$ corresponds to the position central of the local maximum between both defects, while $x_1$ is the position of the global maximum.
 \textbf{d}, Normal mode frequencies of the emergent band, as a function of the number of supercells, calculated with the tight-binding model (circles) and with the transfer matrix method (crosses). 
\textbf{e}, Dispersion relation of the mechanical crystal. The emergent band in {\bf d} lies within the region limited by dispersion relation.
}
\label{figura1co}
\end{figure}

The experimental results given in Figs.~\ref{zzzUltimo31abcd} and~\ref{Figura5} allow us to use the tight-binding approximation, \textit{\`a la} quantum mechanics, to calculate the torsional spectrum of the elastic crystal.
In other words, a mechanical system which can be described by the Anderson model was built.
The basis of the mechanical TB model will be formed by the torsional wave amplitudes $\{\phi_n(x)\}$, localized at the position of the defect, with frequency $f_n$ lying inside a gap of the coupler, associated to each isolated supercell on site $n$ (see Fig.~\ref{figura1co} {\bf a}). 
The torsional wave amplitude $\theta(x)$ of the elastic crystal, as function of position, can be expanded in terms of the basis as
\begin{eqnarray}
\theta (x)=\sum_n A_n \phi_n(x). 
\label{teta}
\end{eqnarray}
Then the tight-binding model for the 1D elastic crystal can be written as 
\begin{eqnarray}
-CA_{n-1}+f_nA_{n}-CA_{n+1}=fA_n, 
\label{ecuacion1resultadosteoricos}
\end{eqnarray}
where, $f_n$, is the resonant frequency of one isolated supercell on site $n$; $C$ is the coupling coefficient between nearest neighbor resonators, which depends also on the properties of the coupler. 
The resonant frequency of the complete crystal will be $f$. 
This is the simplest TB model, since it assumes only nearest-neighbor couplings that are identical for all resonators. Its generalization to more complex situations including second neighbours, as well as 2D and 3D systems, is straightforward. We should note that this model is dynamically equivalent to the quantum-mechanical case, and it differs from the mass-spring tight-binding model (previously considered, \cite{MarcosSoukoulis}), since the latter has $\omega^2$, instead of $f$, being $\omega$ the angular frequency. 
Thus $f$, $f_n$, $\phi_n$ and $C$ in the model given in Eq.~(\ref{ecuacion1resultadosteoricos}) for mechanical waves take the role of the energy, the site energy, the orbital, and the hopping in the quantum TB model, respectively.  
The model represented in Eq.~(\ref{ecuacion1resultadosteoricos}) has two free parameters, the site-frequency  $f_n$ and the mechanical hopping $C$.  Physically  $f_n$  roughly corresponds to the resonant frequency of the isolated resonator whereas $C$ is related to the localization length of the wave amplitude of one resonator within the periodic structure. 
Both values should be determined to generate the TB torsional frequency spectrum for an elastic crystal with an arbitrary number of cells, as those given in Fig.~\ref{zzzUltimo31abcd}. 
The values of $f_n$ and $C$ were obtained from the numerical calculations, made with the transfer matrix method, for a structure with two coupled resonators (Fig.~\ref{figura1co} \textbf{b}). 
In this case the TB model gives two solutions $f_\pm=f_0 \pm C$, thus giving the site-frequency $f_0$ and the mechanical hopping.
Figure~\ref{figura1co}~\textbf{c} gives the ratio $r$ between the total maximum and the central local maximum of the symmetric wave amplitude of Fig.~\ref{figura1co}~\textbf{b}, a measure related to the localization length for the two resonators case, as a function the number of cells of the coupler. 
The level splitting $\Delta f =2 C$ is also given in Fig.~\ref{figura1co}~\textbf{c}. As it can be seen in the same figure, the localization (level splitting) increases (decreases) as the number of cells in the coupler increases and that both quantities roughly show an exponential behavior. One can also notice that the slope of the level splitting is approximately minus twice the slope of the ratio $r$ since the level repulsion can be understood as the overlap of two wave amplitudes of the basis located at different sites. 
%
In Fig.~\ref{figura1co} \textbf{c} the emergent band, appearing around 25.89~kHz is given as function of number of supercells. As it can be seen the results obtained with the tight-binding model agree with those obtained with the transfer matrix method.
It is possible to observe,  that the frequency levels split symmetrically around of the level of frequency $f_0$ in contrast with other elastic crystals studied in the literature \cite{MoralesFloresGutierrezMendezSanchez} whose levels are distributed asymmetrically towards one side of the first frequency level. 
This figure shows that the spectrum of the elastic crystal constructed here is completely analog to a spectrum of an 1D atomic crystal since the resonators interact through the bandgap of the couplers.
When considering an ideal elastic crystal, constructed from an infinite number of supercells (see figure \ref{zzzUltimo31abcd}), it is possible to obtain the dispersion relation of the crystal. 
The wave amplitude  on site $n$ is written as a plane wave, $A_n=e^{-inka}$, where $a$ is the distance between consecutive sites on the elastic crystal and $k$ is the wave number. 
The dispersion relation 
\begin{equation}\label{eqRelacionDispersion}
f= f_0-2C\cos(ka),
\end{equation}
is the same as that of a 1D cristal of tightly bound electrons.
This dispersion relation is shown in Fig.~\ref{figura1co} e); one can notice that the emergent band of the locally periodic system of $n$ supercells, Fig.~\ref{figura1co} d),  is completely contained inside the limits of the dispersion relation, $f_0\pm 2C$, of the ideal elastic crystal.
The group velocity can be obtained directly  from the dispersion relation and it implies that, at the bottom of the band, there are free mechanical quasiparticles analog to the free electrons in an 1D atomic crystal. In the mechanical case the quasiparticles have effective mass $m_\mathrm{eff}=\frac{1}{2 C a^2}$.

In summary: in this article it was shown that it is possible to emulate tightly bound electrons in an atomic crystal using mechanical waves. 
This was possible thanks to the design of an elastic crystal composed of resonators joined with couplers in such a way that the levels associated with the resonators lie within the coupler's bandgap. 
A generalization of resonant acoustic spectroscopy was used to measure the spectra and wave amplitudes of the artificial elastic crystals. 
The results obtained experimentally are well reproduced by quantum tight-binding model.
A complete analogy between the quantities used in the quantum model appear in the mechanical system: the on-site frequency, the mechanical hopping, the mechanical orbitals, the ``frequency'' levels, quasi-particles and an effective mass. 
The envelope of the atomic orbitals and the dispersion relation can also be obtained.
The mechanical systems reported here can be easily generalized to more complex quantum systems in 2D or 3D, such as molecules or graphene among many others.
For obvious reasons, this gives the possibility of analyzing problems that cannot be studied easily in atomic systems due to experimental limitations, with the advantages and challenges of the macroscopic systems used.
The mechanism described here allows to create mechanical materials that have the features of a quantum crystal.  
It also opens  the  door  to  the  study  of some quantum properties using mechanical waves although the results presented here are novel and have their value by themselves. 
The previous statements are supported by the fact that coupling engineering in structured mechanical systems is flexible enough to introduce length deformations that generalize simple periodic configurations. 
The realization of evanescent couplings extends well beyond the mathematical abstraction of Wannier functions \cite{MarzariEtAl,RomanoEtAl} and the limitations of electronic transport in traditional solids.

\section*{Acknnowledgements}

This work was supported by DGAPA-UNAM under project IN109318 and by CONACYT under projects 284096 and AI-S-33920. FRR acknowledges a fellowship from CONACYT.
The authors acknowledge the kind hospitality of Centro Internacional de Ciencias A. C. for group meetings frequently celebrated there and for space to locate the laboratory of waves and materials where part of the experiments were performed. We would like to thank Elisa Guillaumin for invaluable comments.

\section*{Author contributions}
E.S., G.B. and R.A.M.-S. conceived the project. 
F.R.-R. performed the experiment with acoustic resonant spectroscopy, figures 1-3 and carried out the numerical simulations with TB model, with COMSOL and with the transfer matrix method. 
E. F.-O. autimatized the setup and helped with the measurements. 
G.B., F.R.-R. and R.A.M.-S. designed and directed the experiment and wrote the main parts of the text. 
All authors participated in discussions of results and helped improve the manuscript. 

\section*{Competing interests}
The authors declare no competing interests.\\

\section*{Methods}
\subsection{Acoustic resonant spectroscopy}

This technique is used to measure the mechanical vibration spectra for any elastic system and the setup is given in Fig.~\ref{zzzUltimo31abcd}. 
The procedure starts with the generation of a harmonic signal, of frequency $f_0$ in the vector network analyzer (VNA). 
This signal from the VNA is sent to a high-fidelity audio amplifier (Cerwin-Vega 2800 was used) to increase its power.
The output of the amplifier is sent to an acoustic electromagnetic transducer (EMAT) located at the vicinity of the artificial elastic crystals. 
The EMAT, by electromagnetic induction generates mechanical vibrations in the aluminum piece \cite{MoralesGutierrezFlores}. 
A second EMAT detects the mechanical response of the crystal, at other location the beam,  and converts it into a voltage signal. 
This signal is captured by the VNA. 
A workstation is used for the automated storage and subsequent analysis of the data of the different measurements. 
Then $ f_0 $ is changed to $ f_0 + \Delta f $ and the procedure is repeated to obtain an spectrum.
Since the EMATs are non contacting transducers, the position of has to be taken at maximum amplitude; the latter was obtained from the transfer matrix calculations.
The numerical wave amplitudes of Fig.~\ref{figura1co} allow us to identify the locations of the global maximum and minimum of the torsional vibrations to excite and detect spectrum efficiently; they are located at the edges of the resonators.
By moving the EMAT detector along the beam, it is possible measure the wave amplitudes as a function of the position. 

\subsection{Electromagnetic Acoustic transducers (EMATS)}

We will first discuss how these devices excite mechanical waves (See Fig.~\ref{zzzUltimo31abcd}).
A harmonic current is applied to the EMAT coil and the latter produces a magnetic field, also alternating, which induces eddy currents in the metal of the rod. 
The interaction, via Lorentz force, between the magnetic field of the magnet and the eddy currents produce a force on the metal~\cite{MakarovPerezRodriguezYampolskii}. 
The same device can be used as a detector: when the metal surface oscillates close to the EMAT's magnet, the magnetic flux through any loop of the paramagnetic metal will change. 
This, according to Faraday's law, originates an electromotive force in the loops which in turn generates a magnetic field measured by the detector's coil. A deeper explanation of the EMAT operation can be found in Refs.~\cite{FrancoVillafane, Arreola}.

\subsection{COMSOL simulations}
COMSOL Multiphysics was used to calculate the finite element method simulations of Fig.~\ref{Figura5} with the parameters corresponding to aluminum: Young's module $E=68.6$~GPa, Poisson's coefficient $\nu=0.33$ and density $\rho= 2722$~kg~m$^{-3}$. Free boundary conditions were imposed and a symmetrical grid was used.

\subsection{Transfer Matrix}

Lets consider a finite beam along the $z$-axis consisting of $M$ cuboids of square cross section and side $w_i$ with $i=1,2,\dots, M$. 
By definition the transfer matrix relates the amplitudes of the plane waves of $i$-cuboid with those of $(1+1)$-cuboid as
\begin{equation}
\left(
\begin{array}{lcr}
A_{i+1}\\
B_{i+1}\\
\end{array}
\right)
=\frac{1}{2}
\left(
\begin{array}{lcr}
(1+\frac{w_i^4}{W_{i+1}^4})e^{ik(z_i-z_{i-1})}&(1-\frac{w_i^4}{W_{i+1}^4})e^{-ik(z_i-z_{i-1})}\\
(1-\frac{w_i^4}{W_{i+1}^4})e^{ik(z_i-z_{i-1})}&(1+\frac{w_i^4}{W_{i+1}^4})e^{-ik(z_i-z_{i-1})}\\
\end{array}
\right)
\left(
\begin{array}{lcr}
A_{i}\\
B_{i}\\
\end{array}
\right),
 \label{4}
\end{equation}
where the torsion in cuboid $i$, of width $w_i$ ($W_i$) and height $w_i$ ($W_i$) and located between positions $z_{i-1}$ and $z_i$,  is $\phi_i(z)=A_ie^{ik(z-z_{i-1})}+B_ie^{-ik(z-z_{i-1})}$. 
The continuity conditions for the torsion and the moment of torsion at $z = z_i$ are  $\left. \phi_i\right|_{z_i}=\left. \phi_{i+1}\right|_{z_i}$ and $\left. w_i^4 \frac{\partial \phi_i}{\partial z}\right|_{z_i}=\left. W_{i+1}^4 \frac{\partial \phi_{i+1}}{\partial z}\right|_{z_{i+1}}$.
For a bar with square transversal section is given by $c=0.92\sqrt{\frac{G}{\rho}}$, where $G$ is the shear modulus and $\rho$ is the density.

Defining the total transfer matrix as $T=T_{M-1  \rightarrow M}\cdotp \cdotp \cdotp T_{i  \rightarrow i+1} \cdotp \cdotp \cdotp T_{1  \rightarrow 2}$, the amplitudes at the right end of the beam can be written in terms of the amplitudes of the left end as
\begin{equation}
\left(
\begin{array}{lcr}
A_{M}\\
B_{M}\\
\end{array}
\right)
=T\left(
\begin{array}{lcr}
A_{1}\\
B_{1}\\
\end{array}
\right).
 \label{4}
\end{equation}
Finally, using the free-free boundary conditions the normal-mode frequencies of the structured beam are obtained finding the roots of the following equation 
\begin{equation}
    T_{12}+T_{11}e^{ik(L-z_{M-1})}+T_{22}+T_{21}e^{-ik(L-z_{M-1})}=0.
\end{equation}
More details about the transfer matrix method applied to the elastic beams can be found in Ref.~\cite{ArreolaLucas}.

\end{document}